\documentclass[prb,preprint,showpacs,preprintnumbers,amsmath,amssymb]{revtex4}

\usepackage{graphicx,color,psfrag}
\usepackage{dcolumn}
\usepackage{bm}

\begin{document}

\title{Chirality dependence of the radial breathing phonon mode density in single
wall carbon nanotubes}
\author{A. N. Vamivakas$^1$}
\email{nvami@bu.edu}
\author{Y. Yin$^2$}
\author{A. G. Walsh$^2$}%
\author{M. S. \"{U}nl\"{u}$^1$}
\author{B. B. Goldberg$^2$}
\author{A. K. Swan$^1$}
\affiliation{%
$^1$Department of Electrical and Computer Engineering, Boston
  University, 8 St. Mary's St., Boston, Massachusetts 02215, USA\\
$^2$Department of Physics, Boston
  University, 590 Commonwealth Ave., Boston, Massachusetts 02215}%

\date{\today}
\begin{abstract}
A  mass and spring model is used to calculate the phonon mode dispersion for single wall carbon
nanotubes (SWNTs) of
arbitrary chirality.  The calculated dispersions are used to
determine the chirality dependence of the radial breathing phonon mode (RBM)
density.  Van Hove
singularities, usually discussed in the context of the single particle
electronic excitation spectrum, are found in the RBM density of
states with distinct qualitative differences for zig zag, armchair and chiral SWNTs.  The
influence the phonon mode density has on the two phonon
resonant Raman scattering cross-section is discussed.    
\end{abstract}

%
\maketitle


  Understanding how spatial confinement influences both a material's
excitation spectrum and the density of excitation modes is imperative
to accurately modeling the observable properties of the
system.  In particular, there has been much recent interest in the
nanoscale confinement of 
phonons\cite{PhonNano}.
  In one-dimensional (1D)
single wall carbon nanotubes (SWNTs),
there are various methods employed to calculate the phonon mode
dispersion.  Early work, following the successful calculations of the
single electron bandstructure, projected the two-dimensional (2D) graphene phonon
disperion relation to 1D.  It was quickly realized, though,
that zone-folding did not capture all the phonon modes of SWNTs
\cite{Reich}.  
Following some \textit{ab-initio} work \cite{Orde99a,Reic02a}, a
microscopic mass and spring model has been formulated
\cite{Maha04a,Yang06a} to accurately capture the lattice dynamical
properties of SWNTs.  The mass and spring model permits calculation of 
both the phonon dispersion relation and phonon mode density for SWNTs of arbitrary chirality.  

  In the optical characterization of SWNTs, one phonon
 resonant Raman scattering (1phRRS) is employed to measure a tube's electronic
 resonances, investigate various Raman active phonon modes and
 even determine the physical
 diameter and chirality of the tube under study \cite{Reich}.  Though 1phRRS is a
 powerful tool for SWNTs characterization,
 conservation of momentum allows only a single phonon to contribute to
 the measured signal at a particular scattered photon frequency.
In contrast, in two phonon Raman scattering, multiple phonon pairs
originating from the full phonon dispersion curve can
contribute to the scattered signal.
Therefore, the two phonon resonant Raman scattering (2phRRS)
cross-section, unlike the 1phRRS case, is proportional to the phonon
joint density of states (DOS). 
 The 2phRRS
cross-section is nonzero for all scattered
frequencies for which the product of the phonon joint DOS and
 transition matrix element is nonzero.   Although 2phRRS has been discussed
in the context of SWNTs , the discussion has only 
focused on calculating $|W^{2-ph}_{i \rightarrow f}|^2$  and little attention has been paid to the
density of phonon modes \cite{Reich}.  It is clear then, in
2phRRS, an accurate model of both the phonon dispersion and joint DOS is important.



 To calculate the phonon dispersion and DOS we model the nanotube lattice
as a collection of equal mass points, capable of displacing in all
three spatial dimensions.  We use Newton's law $\vec{\mathbf{F}} = -k\vec{\mathbf{x}}= m\ddot{\vec{\mathbf{a}}}$ to understand the
dynamical evolution of each lattice site \cite{Maha04a,Yang06a}.  The governing equation is 

\begin{equation}
\label{DMat}
\Omega^2\vec{Q}=\underline{\mathsf{W}}(\alpha,q;n,m)\vec{Q}
\end{equation}

\noindent where $(n,m)$ index the nanotube, $\alpha=0,1,...,N-1$,
$N=\frac{2(n^2+m^2+nm)}{gcd(2n+m,2m+n)}$ equals the number
of graphene unit cells in the nanotube unit cell, $q$ is the
 1D phonon crystal momentum along
the tube axis,
$\Omega^2=m\omega^2/K_1$ is the rescaled phonon mode frequency where
$K_1$ is the force constant that characterizes the strength of the
various couplings to be described below, 
$\vec{Q} = [Q_{A,\rho} \,\,  Q_{A,\theta} \,\, Q_{A,z}\,\, Q_{B,\rho} \,\,
  Q_{B,\theta} \,\,  Q_{B,z}]$
is the vector of fourier amplitudes for lattice site displacements with $Q_{i,j}$ indicating
the carbon atom at basis site $i$ displaces in the direction $j$ and
the matrix
 dictating the dynamics is

\begin{equation}
\label{Wmat}
\underline{\mathsf{W}} =\Biggl( s_1M_{1stNN} + s_2M_{2ndNN}+ s_3M_{rbb}\Biggr).
\end{equation}

\noindent  In $\underline{\mathsf{W}}$, $M_{1stNN}$ is a 6x6 matrix characterizing the
influence first nearest
neighbor carbons atoms have in displacing a specific basis atom, $M_{2ndNN}$
is a 6x6 matrix characterizing the influence second nearest neighbor
carbon atoms have in displacing a specific basis atom and
$M_{rbb}$ is a 6x6 matrix characterizing the restoring force on a given
basis site resulting from bending the bond between neighboring
carbon atoms.   The above dynamics allow each carbon atom to displace in the
$\rho$, $\theta$, and $z$ directions.   In Eq. \eqref{Wmat}, $s_i$ determines
the relative strength of the given coupling with respect to the
coupling characterized by the spring
constant $K_1$.  Specifically, $s_i = \frac{K_i}{K_1}$ where $K_i$ is
the spring constant characterizing the $i^{th}$ restoring force.
The parameter $\alpha$ serves as a band index for the
phonons, where the number of phonon bands for a
given nanotube is equal to three times the number of carbon atoms in the nanotube
unit cell.  For fixed $\alpha$, the phonon mode dispersion relation is calculated
by diagonalizing Eq. \eqref{Wmat} as a
function of $q$ as $q$ ranges throughout the nanotube's first
Brillouin zone.  

In addition to determining the dispersion relation, it is also
possible to calculate the 1D phonon density of states.
Generally, the
density of phonon modes is an important quantity when calculating
transition rates for scattering processes mediated by phonons.
We are particularly interested in the two phonon Raman scattering process.
The two phonon mediated Raman scattering cross-section, assuming
free electrons and holes, is \cite{CardScatt} 


\begin{equation}
\label{TPRRS}
\frac{d\sigma^{2-ph}_{RRS}}{d\Omega d\omega_s} =
C\rho^{(2-ph)}_{DOS}(2\Omega_p)\cdot
|W^{2-ph}_{i\rightarrow f}
(\omega_l,\vec{e_l};\omega_s=\omega_l-2\Omega_p,\vec{e_s})|^2
\end{equation}

\noindent where we have assumed Stokes scattering, $C$ is a constant, 
$\rho^{(2-ph)}_{DOS}(2\Omega_p)$ is the two phonon density of states 
and $|W^{2-ph}_{i \rightarrow f}|^2$ is the transition probability from
initial system state $i$, with a single pump photon, to a final state
$f$, with a single scattered photon and a pair of degenerate single
phonons with opposite crystal momentum.  We note, in Eq. \eqref{TPRRS}
our focus is on final states that contain two
degenerate phonons of frequency $\Omega_p$.  In 2phRRS, it is also
possible to have scattered photons at $\omega_s = \omega_l -
(\Omega_{p1}\pm\Omega_{p2})$, the sum or difference frequency of
two nondegenerate phonons.  In a 1-dimensional system, our expectation is the
reduction of phonon phase space will hamper the likelihood of
simultaneously satisfying both energy and momentum conservation in the
sum or difference frequency generation process so we have ignored this
possibility in Eq. \eqref{TPRRS}.

The utility in focusing on the degenerate phonon generation, is that we can
approximate $\rho^{(2-ph)}_{DOS}(2\Omega_p)$ by
$\rho^{(1-ph)}_{DOS}(\Omega_p)$ \cite{CardScatt}.  With the
single particle density of states defined as \cite{Kittel}

\begin{equation}
\label{DOS}
\rho_{DOS}(\Omega)d\Omega \propto \frac{1}{|\frac{d\Omega}{dq}|}d\Omega
\end{equation}  

\noindent we can evaluate $\rho^{(1-ph)}_{DOS}(\Omega_p)$ for a given
phonon band with knowledge of $\Omega_p(q;\alpha)$.  In degenerate 2phRRS,
$\rho^{(1-ph)}_{DOS}(\Omega_p)$ serves as an indicator of how many
phonon pairs contribute to a specific scattered photon frequency.
Specifically, the  regions of zero slope in the phonon
dispersion result in Van Hove singularities \cite{Hove54a} that
enhance the 2phRRS cross-section.  



   To illustrate the chirality dependence of the RBM single phonon density
   of states, we study nanotube families $2n+m=22$ and
   $2n+m=23$ as well as subsets of both armchair and zig zag tubes.  Our interest in
   family 22 stems from recent experimental work in 2phRRS from single SWNTs
   \cite{Swan05a}.  In addition, family 22 and 23 yields a set of 
  semiconducting SWNTs with chiral angles ranging from
   $0^{\circ}$ to $30^{\circ}$.

\begin{figure}[t]
\begin{center}
\includegraphics[width=5.34in]{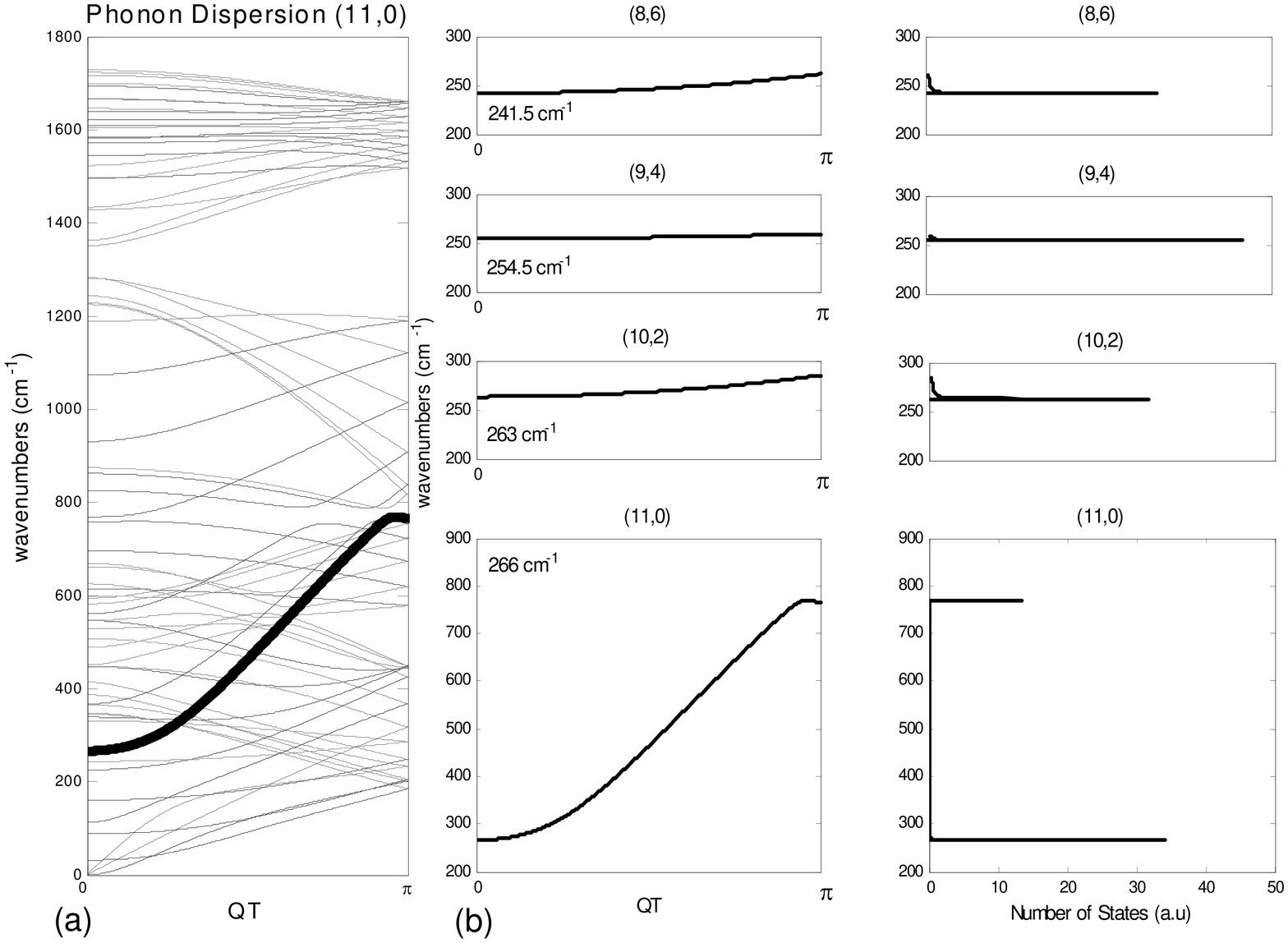}
\end{center}
\caption{(a) The phonon dispersion relation for an (11,0) nanotube.  The bold line is the RBM
 dispersion curve. (b)  The  RBM dispersion and RBM DOS for family 22
  SWNTs.  The number in the dispersion figures is the zone-center RBM frequency.  In all figures,  T is
  the magnitude of
  the primitive vector along the tube axis.}
\label{Fig1}
\end{figure}

Before studying the radial breathing mode (RBM) DOS, we illustrate a typical
phonon dispersion relation calculated from Eq. \eqref{Wmat}.  Figure \ref{Fig1}(a) is the
full phonon dispersion relation for an (11,0) nanotube (family 22).
The number of carbon atoms in the nanotube unit cell is 44 which
results in $3*44=132$ phonon modes. Also, the length
of the primitive lattice vector along the tube axis is 3 times the
carbon-carbon distance on the graphene lattice.  To fix the frequency
axis, we have followed Mahan\cite{Maha04a} and set $\Omega=\sqrt{3}$ equal
to $\omega = 1600 \,\text{cm}^{-1}$ (this fixes the value of the free
parameter $K_1$).  We also assumed
in Eq. \eqref{Wmat} that $s_1=1$,
$s_2=0.06$ and $s_3=0.024$ \cite{Maha04a}.
The calculated zone center frequency for the RBM of the (11,0) tube is
$266\, \text{cm}^{-1}$.  The calculated value is in near perfect agreement with
recently reported RBM-diameter relation of $\omega_{RBM} =
\frac{215}{D}+18$ where $D$ is the nanotube diameter\cite{Reic05a},
which yields a zone center RBM
frequency of $267.4\,\text{cm}^{-1}$ for the (11,0) nanotube. 

Now we investigate the RBM DOS chirality dependence. In Fig.
\ref{Fig1}(b), we plot the RBM dispersion
curve and RBM
DOS for SWNTs of family 22.  With the exception of the
tube (11,0),
 which will be discussed when we focus on the zig
zag tubes,  the other
tubes comprising family 22 have RBMs with mode densities 
concentrated entirely at the zone center.  In addition, the chiral tubes
of family 22 have RBMs that exhibit very low dispersion across the
nanotube Brillouin zone.  The Van Hove singularity in the RBM
mode density of family 22's chiral tubes results in a 2phRRS
cross-section that is entirely mediated by the zone center RBM.  In
this case, even though the momentum conservation has been relaxed in
the two phonon scattering process, the phonon mode density prohibits
phonons of appreciable crystal momentum from contributing to the 2phRRS
cross-section.  Family 23 exhibits qualitatively identical
behaviour to family 22.

\begin{figure}[t]
\begin{center}
\includegraphics[width=5.34in]{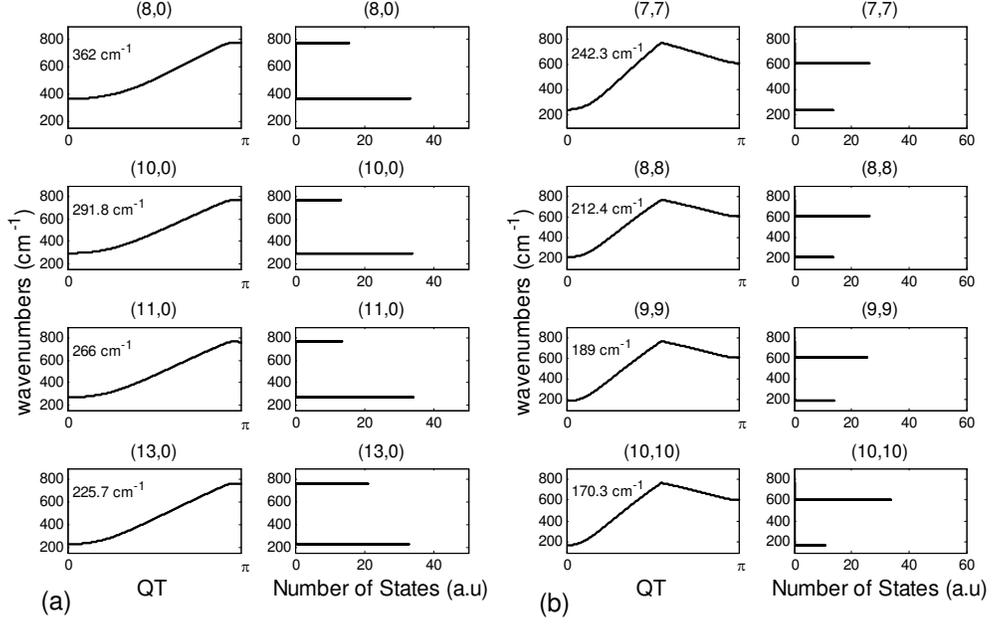}
\end{center}
\caption{(a)  The RBM dispersion and RBM DOS for zig zag SWNTs
  (8,0), (10,0), (11,0) and (13,0). (b)  The RBM dispersion and RBM
  DOS for armchair SWNTs
  (7,7), (8,8), (9,9) and (10,10).   The number in the dispersion
  figures is the zone-center RBM frequency.
  In all figures, T is
  the magnitude of
  the primitive vector along the tube axis.}
\label{Fig2}
\end{figure}

Figure \ref{Fig2} contains plots of the RBM dispersion, and its associated DOS, for
semiconducting zig zag tubes (8,0), (10,0), (11,0) and (13,0) where we note that the RBM bandwidth decreases with decreasing tube
diameter.  There
are two striking features in Fig. \ref{Fig2}(a) when compared to the chiral
tubes of family 22. First, the bandwidth of the RBM is considerably
larger than the chiral tubes, nearly 2 orders of magnitude in some
cases.  Second, zig zag tubes exhibit two sharp Van Hove singularities in
their RBM mode density.  If we define the zone boundary of the
Brillouin zone as $q_{max}$, we find that only wavevectors out to
$0.03*q_{max}$ contribute to the zone center singularity.  In 2phRRS, when the incoming laser frequency is
commensurate with an electronic resonance, the two singularities
in the RBM mode density should result in two peaks in the scattering
cross-section, with a higher intensity peak associated with a
frequency that is twice the zone center RBM frequency.  

Finally, we examine the RBM dispersion and DOS, plotted
in Fig. \ref{Fig2}(b), for the
(metallic) armchair tubes (7,7), (8,8), (9,9) and (10,10).  In
comparison to the previous zig zag tubes, the strongest
Van Hove singularity for the RBM has shifted from zone center to the
vicinity of the
zone boundary.  Approximately $0.01*q_{max}$ wavevectors contribute to the
zone boundary singularity.  The
previous observation has an important consequence in
 2phRRS.
Not only will we observe a double peak in the scattering cross-section
when the incoming laser frequency is tuned to an electronic resonance, but the more intense peak will be
associated with a frequency that is equal to twice the zone boundary
RBM phonon frequency.                    

     In summary, we have used a microscopic model of the lattice
     dynamical properties of SWNTs to determine the chirality
     dependence of the RBM mode density.  The importance of the phonon
     mode density for 2phRRS was discussed and illustrated.  For chiral
     tubes of family 22 and 23 (with the exception of the (11,0) tube), we
     found narrow band RBMs with 
  Van Hove singularities
     restricted to the zone center.  In 2phRRS, we expect to see
     spectral features at scattered frequencies that are exactly 2 times
     the zone center RBM phonon frequency as determined by 1phRRS.
     
     In contrast to the chiral tubes, the set of achiral SWNTs studied 
     exhibited
     large bandwidth RBMs with 2 Van Hove singularities in their DOS.  
  The zone center Van Hove singularity dominated for zig
     zag tubes, whereas the zone boundary singularity dominated for
     armchair SWNTs.  The appearance of two Van Hove singularities in
     achiral tube's mode density allows for the possibility of a 2
     peak structure in the 2phRRS cross-section at fixed laser frequency.

\begin{acknowledgments}
This work was supported by Air Force Office of Scientific Research
under Grant No. MURI F-49620-03-1-0379, by NSF under Grant No. NIRT
ECS-0210752 and a Boston University SPRInG grant.  The authors would
like to thank Gun Sang Jeon for discussions on his microscopic
model of achiral nanotube phonons\cite{Maha04a}.
\end{acknowledgments}


\begin{thebibliography}{11}
\expandafter\ifx\csname natexlab\endcsname\relax\def\natexlab#1{#1}\fi
\expandafter\ifx\csname bibnamefont\endcsname\relax
  \def\bibnamefont#1{#1}\fi
\expandafter\ifx\csname bibfnamefont\endcsname\relax
  \def\bibfnamefont#1{#1}\fi
\expandafter\ifx\csname citenamefont\endcsname\relax
  \def\citenamefont#1{#1}\fi
\expandafter\ifx\csname url\endcsname\relax
  \def\url#1{\texttt{#1}}\fi
\expandafter\ifx\csname urlprefix\endcsname\relax\def\urlprefix{URL }\fi
\providecommand{\bibinfo}[2]{#2}
\providecommand{\eprint}[2][]{\url{#2}}

\bibitem[{\citenamefont{Stroscio and Dutta}(2005)}]{PhonNano}
\bibinfo{author}{\bibfnamefont{M.~A.} \bibnamefont{Stroscio}} \bibnamefont{and}
  \bibinfo{author}{\bibfnamefont{M.}~\bibnamefont{Dutta}},
  \emph{\bibinfo{title}{Phonons in Nanostructures}}
  (\bibinfo{publisher}{Cambridge University Press}, \bibinfo{year}{2005}).

\bibitem[{\citenamefont{Reich et~al.}(2004)\citenamefont{Reich, Thomsen, and
  Maultzsch}}]{Reich}
\bibinfo{author}{\bibfnamefont{S.}~\bibnamefont{Reich}},
  \bibinfo{author}{\bibfnamefont{C.}~\bibnamefont{Thomsen}}, \bibnamefont{and}
  \bibinfo{author}{\bibfnamefont{J.}~\bibnamefont{Maultzsch}},
  \emph{\bibinfo{title}{Carbon Nanotubes}} (\bibinfo{publisher}{Wiley, VCH},
  \bibinfo{year}{2004}).

\bibitem[{\citenamefont{S\'{a}nchez-Portal
  et~al.}(1999)\citenamefont{S\'{a}nchez-Portal, Artacho, Soler, Rubio, and
  Ordej\'{o}n}}]{Orde99a}
\bibinfo{author}{\bibfnamefont{D.}~\bibnamefont{S\'{a}nchez-Portal}},
  \bibinfo{author}{\bibfnamefont{E.}~\bibnamefont{Artacho}},
  \bibinfo{author}{\bibfnamefont{J.~M.} \bibnamefont{Soler}},
  \bibinfo{author}{\bibfnamefont{A.}~\bibnamefont{Rubio}}, \bibnamefont{and}
  \bibinfo{author}{\bibfnamefont{P.}~\bibnamefont{Ordej\'{o}n}},
  \bibinfo{journal}{Phys. Rev. B} \textbf{\bibinfo{volume}{59}},
  \bibinfo{pages}{12678} (\bibinfo{year}{1999}).

\bibitem[{\citenamefont{Maultzsch et~al.}(2002)\citenamefont{Maultzsch, Reich,
  Thomsen, Dobard\u{z}i\'{c}, Milo\u{s}evi\'{c}, and
  Damnjanovi\'{c}}}]{Reic02a}
\bibinfo{author}{\bibfnamefont{J.}~\bibnamefont{Maultzsch}},
  \bibinfo{author}{\bibfnamefont{S.}~\bibnamefont{Reich}},
  \bibinfo{author}{\bibfnamefont{C.}~\bibnamefont{Thomsen}},
  \bibinfo{author}{\bibfnamefont{E.}~\bibnamefont{Dobard\u{z}i\'{c}}},
  \bibinfo{author}{\bibfnamefont{I.}~\bibnamefont{Milo\u{s}evi\'{c}}},
  \bibnamefont{and}
  \bibinfo{author}{\bibfnamefont{M.}~\bibnamefont{Damnjanovi\'{c}}},
  \bibinfo{journal}{Solid State Commun.} \textbf{\bibinfo{volume}{121}},
  \bibinfo{pages}{471} (\bibinfo{year}{2002}).

\bibitem[{\citenamefont{Mahan and Jeon}(2004)}]{Maha04a}
\bibinfo{author}{\bibfnamefont{G.~D.} \bibnamefont{Mahan}} \bibnamefont{and}
  \bibinfo{author}{\bibfnamefont{G.~S.} \bibnamefont{Jeon}},
  \bibinfo{journal}{Phys. Rev. B} \textbf{\bibinfo{volume}{70}},
  \bibinfo{pages}{075405} (\bibinfo{year}{2004}).

\bibitem[{Yan()}]{Yang06a}
\bibinfo{note}{W. Mu, A. N. Vamivakas, and Z. can Ou-Yang, {\it to be
  submitted} Phys. Rev. B}.

\bibitem[{\citenamefont{Cardona and G{\"{u}}ntherodt}(1982)}]{CardScatt}
\bibinfo{editor}{\bibfnamefont{M.}~\bibnamefont{Cardona}} \bibnamefont{and}
  \bibinfo{editor}{\bibfnamefont{G.}~\bibnamefont{G{\"{u}}ntherodt}}, eds.,
  \emph{\bibinfo{title}{Light Scattering in Solids II}}
  (\bibinfo{publisher}{Springer, Heidelberg}, \bibinfo{year}{1982}),
  vol.~\bibinfo{volume}{50} of \emph{\bibinfo{series}{Topics in Applied
  Physics}}, chap.~\bibinfo{chapter}{2}, p.~\bibinfo{pages}{19}.

\bibitem[{\citenamefont{Kittel}(1995)}]{Kittel}
\bibinfo{author}{\bibfnamefont{C.}~\bibnamefont{Kittel}},
  \emph{\bibinfo{title}{Introduction to Solid State Physics}}
  (\bibinfo{publisher}{Wiley}, \bibinfo{year}{1995}), \bibinfo{edition}{seventh
  edition} ed.

\bibitem[{\citenamefont{Hove}(1953)}]{Hove54a}
\bibinfo{author}{\bibfnamefont{L.~V.} \bibnamefont{Hove}},
  \bibinfo{journal}{Phys. Rev.} \textbf{\bibinfo{volume}{89}},
  \bibinfo{pages}{1189} (\bibinfo{year}{1953}).

\bibitem[{\citenamefont{Yin et~al.}()\citenamefont{Yin, Cronin, Walsh,
  Stolyarov, Tinkham, Vamivakas, Basca, {\"{U}}nl{\"{u}}, Goldberg, and
  Swan}}]{Swan05a}
\bibinfo{author}{\bibfnamefont{Y.}~\bibnamefont{Yin}},
  \bibinfo{author}{\bibfnamefont{S.}~\bibnamefont{Cronin}},
  \bibinfo{author}{\bibfnamefont{A.}~\bibnamefont{Walsh}},
  \bibinfo{author}{\bibfnamefont{A.}~\bibnamefont{Stolyarov}},
  \bibinfo{author}{\bibfnamefont{M.}~\bibnamefont{Tinkham}},
  \bibinfo{author}{\bibfnamefont{A.~N.} \bibnamefont{Vamivakas}},
  \bibinfo{author}{\bibfnamefont{W.}~\bibnamefont{Basca}},
  \bibinfo{author}{\bibfnamefont{M.~S.} \bibnamefont{{\"{U}}nl{\"{u}}}},
  \bibinfo{author}{\bibfnamefont{B.~B.} \bibnamefont{Goldberg}},
  \bibnamefont{and} \bibinfo{author}{\bibfnamefont{A.~K.} \bibnamefont{Swan}},
  \bibinfo{note}{accepted for publication in \textit{IEEE J. Sel. Top. Quantum
  Electron..}}

\bibitem[{\citenamefont{Maultzsch et~al.}(2005)\citenamefont{Maultzsch, Telg,
  Reich, and Thomsen}}]{Reic05a}
\bibinfo{author}{\bibfnamefont{J.}~\bibnamefont{Maultzsch}},
  \bibinfo{author}{\bibfnamefont{H.}~\bibnamefont{Telg}},
  \bibinfo{author}{\bibfnamefont{S.}~\bibnamefont{Reich}}, \bibnamefont{and}
  \bibinfo{author}{\bibfnamefont{C.}~\bibnamefont{Thomsen}},
  \bibinfo{journal}{Phys. Rev. B} \textbf{\bibinfo{volume}{72}},
  \bibinfo{pages}{205438} (\bibinfo{year}{2005}).

\end{thebibliography}

\newpage



\end{document}